\documentclass[%
 reprint,
superscriptaddress,
 amsmath,amssymb,
 aps,
prd,
longbibliography,
]{revtex4-1}

\usepackage{graphicx}% Include figure files
\usepackage{dcolumn}% Align table columns on decimal point
\usepackage{bm}% bold math
\usepackage{hyperref}% add hypertext capabilities
\usepackage{color}
\usepackage[mathlines]{lineno}% Enable numbering of text and display math
\usepackage{tabularx}
\usepackage{subfigure}
\usepackage[normalem]{ulem}
\usepackage[export]{adjustbox}
\usepackage{url}
\hypersetup{
    pdfnewwindow=true,    % links in new window
    colorlinks=true,      % false: boxed links; true: colored links
    linkcolor=blue,       % color of internal links
    citecolor=blue,       % color of links to bibliography
    filecolor=blue,       % color of file links
    urlcolor=blue         % color of external links
}
\usepackage{orcidlink}

\newcolumntype{Y}{>{\centering\arraybackslash}X}

\usepackage{natbib}
\def\oneG{$\left | B(-1250) \right | = 1\,{\rm G}$}
\def\zeroG{$\left | B(<0) \right | = 0$}

\begin{document}

\author{Kenny C.Y. Ng \orcidlink{0000-0001-8016-2170}}
\email{kcyng@cuhk.edu.hk}
%\thanks{\scriptsize \!\! \href{http://orcid.org/0000-0001-8016-2170}{orcid.org/0000-0001-8016-2170}}
\affiliation{Department of Physics, The Chinese University of Hong Kong, Shatin, New Territories, Hong Kong, China}

\author{Andrew Hillier \orcidlink{0000-0002-0851-5362}}
\email{A.S.Hillier@exeter.ac.uk}
%\thanks{\scriptsize \!\! \href{http://orcid.org/0000-0001-8016-2170}{orcid.org/0000-0001-8016-2170}}
\affiliation{Department of Mathematics and Statistics, University of Exeter, Exeter, EX4 4QF UK}

\author{Shin'ichiro Ando \orcidlink{0000-0001-6231-7693}}
\email{s.ando@uva.nl}
%\thanks{\scriptsize \!\! \href{http://orcid.org/0000-0001-8016-2170}{orcid.org/0000-0001-8016-2170}}
\affiliation{GRAPPA Institute, University of Amsterdam, 1098 XH Amsterdam, The Netherlands}
\affiliation{Kavli Institute for the Physics and Mathematics of the Universe, University of Tokyo, Chiba 277-8583, Japan}

\date{ May 27, 2024}

\title{TeV Solar Gamma Rays as a probe for the Solar Internetwork Magnetic Fields}

% ===== Abstract ===== #
\begin{abstract}
The magnetic fields that emerge from beneath the solar surface and permeate the solar atmosphere are the key drivers of space weather and, thus, understanding them is important to human society.  Direct observations, used to measure magnetic fields, can only probe the magnetic fields in the photosphere and above, far from the regions the magnetic fields are being enhanced by the solar dynamo.  
Solar gamma rays produced by cosmic rays interacting with the solar atmosphere have been detected from GeV to TeV energy range, and revealed that they are significantly affected by solar magnetic fields. However, much of the observations are yet to be explained by a physical model. 
Using a semi-analytic model, we show that magnetic fields at and below the photosphere with a large horizontal component could explain the $\sim$1 TeV solar gamma rays observed by HAWC.  This could allow high-energy solar gamma rays to be a novel probe for magnetic fields below the photosphere.  
\end{abstract}

% \pacs{Valid PACS appear here}
\maketitle

% ====== INTRODUCTION ====== % 
\section{Introduction}

It has long been theorised that cosmic rays impacting the Sun would produce high-energy gamma rays through hadronic interactions~\cite{Seckel:1991ffa}~(hereafter SSG). The pioneering work by SSG considered cosmic-ray reflection by magnetic flux tubes, which significantly enhanced the solar gamma-ray flux compared to the case where the magnetic fields are ignored~\cite{Zhou:2016ljf}.  

Solar atmospheric gamma rays were first detected by EGRET~\cite{Orlando:2008uk}, but precision observation was only made possible by the Fermi space gamma-ray telescope. Ref.~\cite{Abdo:2011xn} showed that the observed flux is much higher than the SSG prediction~\cite{Seckel:1991ffa} in the 0.1--10\,GeV energy range.  Later, Refs.~\cite{Ng:2015gya, Tang:2018wqp, Linden:2018exo, Linden:2020lvz} confirmed this up to around 100\,GeV.  Furthermore, solar gamma rays also exhibit a hard spectrum, large time variation that anticorrelates with solar activity~\cite{Ng:2015gya, Tang:2018wqp, Linden:2020lvz}, a spectral dip in the energy spectrum~\cite{Tang:2018wqp}, and nontrivial time-dependent gamma-ray morphology~\cite{Linden:2020lvz, Arsioli:2024scu}. 
Recently, solar gamma rays were detected for the first time by HAWC~\cite{HAWC:2022khj} in the TeV range with a soft spectral index of about $-3.6$.  

In terms of modeling efforts, Refs.~\cite{Mazziotta:2020uey, Li:2020gch} implemented the PFSS~\cite{Schatten:1969, Altschuler:1969SoPh, Hoeksema:1984PhDT, Wang:1992ApJ} corona magnetic field model in particle propagation and interaction simulation suites to compute the solar gamma-ray flux. While both calculations showed that corona magnetic fields can significantly affect gamma-ray production and could plausibly account for most of the observed flux below 10\,GeV, the observed flux above 100\,GeV remains unexplained.  Ref.~\cite{Gutierrez:2019fna, Gutierrez:2022mor} connected the cosmic-ray Sun shadow and the cosmic-ray induced solar emissions. Ref.~\cite{Banik:2023shc} hypothesized that the Sun could be a TeVatron to explain the high-energy emission. Using 3D MHD simulations, Ref.~\cite{Hudson2020MNRAS.491.4852H} suggested a pressure-magnetic field relation for computing the cosmic-ray interaction in the solar atmosphere.  
Most recently, Ref.~\cite{Li:2023twp} considered magnetic flux tubes and flux sheets, showing an impressive agreement in flux and spectral shape with data from $\sim 10$\,GeV to $\sim 1$\,TeV, though the flux at TeV falls short by a factor of a few under the chosen set of fixed parameters.

Despite all these efforts, there is a lack of concrete theoretical understanding on the production of solar gamma rays. Given the complexity of solar magnetic fields and the vast cosmic-ray energy scale involved (from sub-GeV to beyond TeV), a complete description of solar gamma-ray production may seem daunting without investing significant effort into detailed simulations and modeling. However, as we argue in this work, given a set of reasonable assumptions, we can identify the dominant effect responsible for the gamma-ray production at the high-energy limit. In particular, we show that horizontal sub-photospheric internetwork magnetic fields are likely responsible for the observed $\sim$TeV solar gamma rays.

\section{The need for strong fields}

In this work we exclusively discuss steady solar gamma-ray emission; transient emission, such as those associated with solar flares~\cite{Ramaty1975SSRv...18..341R, Murphy1987ApJS...63..721M, Kozlovsky2002ApJS..141..523K, Omodei:2018uni, Share:2018kqt}, has only been observed up to a few GeV~\cite{Fermi-LAT:2013vao_flares} and are all removed for Fermi solar gamma-ray analyses~(E.g., Refs~\cite{Ng:2015gya, Tang:2018wqp, Linden:2018exo, Linden:2020lvz}.)
There are also leptonic components, such as inverse Compton~\cite{Orlando:2006zs, Moskalenko:2006ta, Orlando:2008uk, Zhou:2016ljf, Orlando:2020ezh, Lai:2022qif, Yang:2023res} and synchrotron~\cite{Petrosian:2022bxr,Orlando:2022xsm} emission by cosmic-ray electrons interacting with Sunlight or solar magnetic fields, but they have completely different morphology and thus can be easily distinguished from the hadronic disk component.

Solar atmospheric gamma rays (also called the disk component) are produced dominantly through hadronic channels, when secondary particles (such as neutral pions) decay into gamma rays. Typically, the produced gamma rays have energy about 10\% of that of the primary cosmic rays, $E_{\gamma} \sim 0.1 E_{p}$. 
The fact that HAWC observed TeV solar gamma rays implies that  cosmic rays up to 10\,TeV must be strongly affected by solar magnetic fields. Considering the Larmor radius~($L$) versus energy, 
\begin{equation}
    \frac{E}{20\, \rm TeV} \simeq \frac{L}{R_{\odot}} \frac{B}{\rm G} \simeq \frac{L}{10^3\,\rm km} \frac{B}{\rm kG} \, ,
\end{equation}
one can see that at least 1\,kG field strength is needed in order to affect $\sim$10\,TeV cosmic rays while confining the physical scale to be around $10^{3}$\,km. 

\begin{figure}[t]
\centering
\includegraphics[width=\columnwidth]{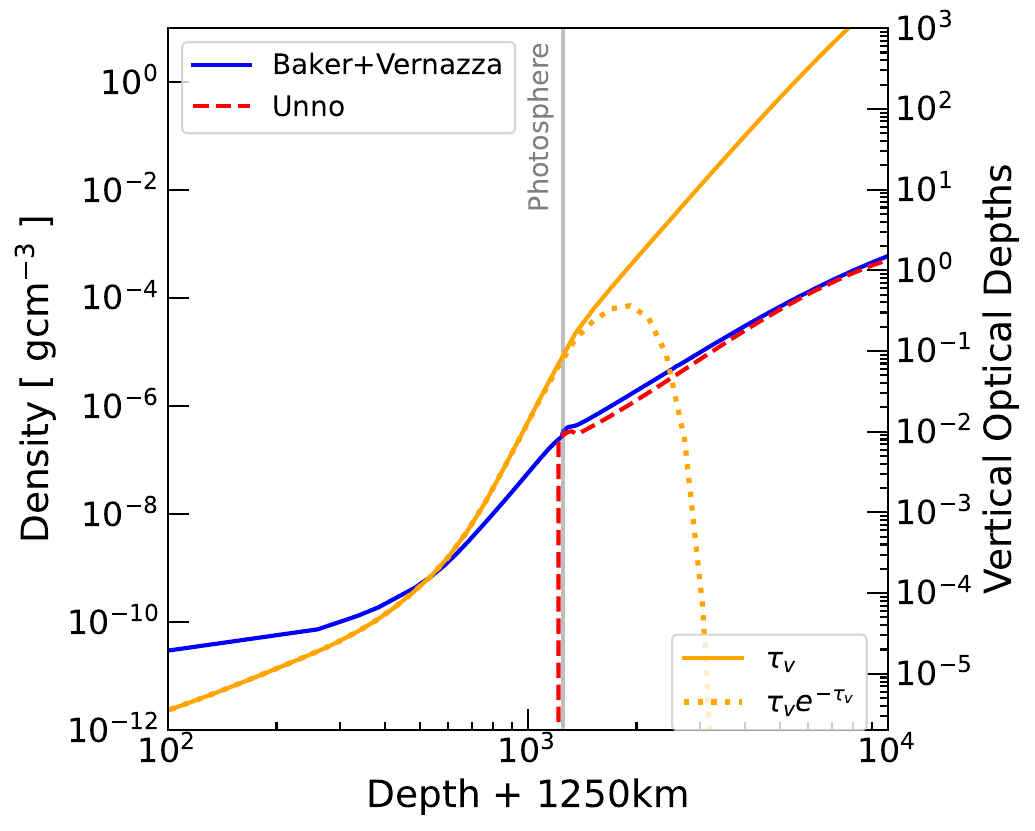}
\caption{Solar density model from Baker~\cite{Baker:1966} and Vernazza~\cite{1973ApJ...184..605V} (blue solid line). We note that these two models and the density solution by Ref.~\cite{Unno:1985PASJ} (red dashed line) overlaps just below the photosphere and agree excellently with each other.  In orange lines, we also show the vertical optical depths, showing that a few hundred km below the photosphere is the most optimal depth for producing solar gamma rays (see text for details).   }
\label{fig:density}
\end{figure}

To give some context for this $10^{3}\,{\rm km}$ length scale, Figure~\ref{fig:density} shows the matter density as a function of {\emph{depth}}~\cite{Baker:1966, 1973ApJ...184..605V}, defined to be zero at the photosphere and increases towards the center of the Sun.  
Given the density profile, one can estimate the gamma-ray production efficiency by computing the vertical optical depth toward the center of the Sun $\tau_{v} = \int \frac{\rho}{m_{p}} \sigma d r$, where $\rho$ and $m_{p}$ are the matter density and proton mass, respectively.
In this work, we take the proton-proton cross section, $\sigma$, to be about 31\,mb~\cite{Kelner:2006tc}, and we ignore the modest increase of the cross section toward high energy. 
In addition, as depth increases, it is increasingly more difficult for the gamma rays to escape the Sun. The combined gamma-ray production and escape efficiency can then be approximated by $\tau_{v}e^{-\tau_{v}}$, as the exponential factor approximates the parent particle's attenuation and the escape probability of the produced gamma rays.  This approximation is justified as the photon radiation length~\cite{Workman:2022ynf} is only slightly longer than the proton interaction length.  Both $\tau_{v}$ and $\tau_{v}e^{-\tau_{v}}$ are shown with the right axis of Figure.~\ref{fig:density}. 
We can see that the $10^2$--$10^{3}$\,km depth scale mentioned above corresponds to the peak of the $\tau_{v}e^{-\tau_{v}}$, which highlights the typical depth required for maximal gamma-ray production efficiency. 
Incidentally, this length scale is also approximately the size of a granule, the convective cell at the solar surface.

%%%%%%%%%%%%%%%%%%%%%%%%%%%%%%%%%%%%%%%% PLOTS
%%%%%%%%%%%%%%%%%%%%%%%%%%%%%%%%%%%%%%%%

One potential site for finding the 1\,kG fields is sunspots.  However, at a given time, sunspots only occupy a small fraction of the solar surface, thus unlikely to affect solar gamma-ray production at a great capacity.
At much smaller scales, kG fields can form in the quiet Sun in intergranular lanes. 
Recently, the effect of the vertical network fields in flux tubes and intergranular lanes were investigated in Ref.~\cite{Li:2023twp} by assuming that cosmic rays follow the open field lines high in the atmosphere into these structures and considering the cosmic rays reflected by the magnetic bottle effect. With a fixed choice of parameters, the flux and the spectral shape agrees well with the data, though it falls below the HAWC observation at TeV by a factor of a few. This indicates that network fields are likely important for the gamma-ray production for a wide energy range. 

In this work, we look at a completely different magnetic field configuration by turning our eyes to another potential source of kG fields: \emph{horizontal} internetwork fields below the photosphere. 

\section{The role of solar internetwork magnetic fields}

The atmosphere of the Sun (photosphere, chromosphere, and Corona) shows a diverse range of magnetic phenomena~\cite{Wedemeyer-Bohm:2008tpk,Wiegelmann2014, Solanki2006RPPh...69..563S}. In the photosphere, it can be divided into active and \emph{Quiet} regions.  Active regions famously host strong magnetic fields features, e.g., in Sunspots.  The quiet regions, though once thought to be non-magnetic, is now known to be fully magnetic~\cite{Almeida:2011talk} (the so-called Quiet Revolution), and could even dominate the magnetic energy of the Sun~\cite{TrujilloBueno:2004re}.

The magnetic fields in the quiet regions come from network and internetwork regions (see Ref.~\cite{Rubio2019LRSP, Rempel2023SSRv} for reviews).
The network regions outline the shape of the granular and supergranular cells, while the internetwork regions represent the cell interiors.  
The internetwork fields, which have typical field strength of ${\cal O}(100)$\,G at the solar surface, have a large horizontal component~\cite{TrujilloBueno:2004re, Lites1996ApJ...460.1019L, Suarez:2007pp, Lites2008ApJ...672.1237L, Rubio2019LRSP} and are suggested to be caused by local dynamo from the ubiquitous convective motion~\cite{Schuessler:2008si, Steiner2008ApJ...680L..85S, Steiner2010ASSP...19..166S, Rempel2023SSRv}.
Given the large coverage of the internetwork regions and that the convective motions below the surface could produce similar fields (but at strengths determined by the local energy equipartition between kinetic and magnetic energy), we consider their role in gamma-ray production. 

For solar gamma-ray production, horizontal fields are likely to be \emph{efficient}, as cosmic rays can be reflected as long as they stay in the coherent field region, then observable gamma rays can be produced when cosmic rays interact after they are reflected. 

As we limit our attention to the TeV gamma rays, we can also safely ignore the processes that affect mostly lower energy cosmic rays or are expected to be subdominant. For example,   
the interplanetary magnetic fields affect how cosmic rays propagate from interstellar space to the surface of the Sun~\cite{Potgieter:2013pdj}, but only affect gamma rays around and below 10 GeV~\cite{Abdo:2011xn, Li:2022zio}.  
Typical corona magnetic fields~\cite{Wang:1992ApJ}, as shown in Ref.~\cite{Mazziotta:2020uey, Li:2020gch}, are not strong enough to significantly affect $>$TeV cosmic rays.  

\subsection {A toy model of the horizontal internetwork fields}
We use a simplified model to investigate the effect of horizontal internetwork magnetic fields on solar gamma-ray production. 
At the solar surface, the typical ${\cal O}(100)$\,G horizontal field strength~\cite{TrujilloBueno:2004re, Lites1996ApJ...460.1019L, Suarez:2007pp, Lites2008ApJ...672.1237L, Rubio2019LRSP} is comparable to the kinetic energy density of the fluid. For reference, with density around $3\times 10^{-7}\,{\rm g/cm^{3}}$ and flow speed around $3\,{\rm km/s}$, the equiparition field is around 580\,G. 
We thus model horizontal field strength below the photosphere following this assumption,
\begin{equation}\label{eq:bmodel}
    |\vec B(r)| = f \sqrt{ 4 \pi \rho(r) x(r)^{2} } \, ,
\end{equation}
where $\rho$ is the density, $r$ is the radial distance, $x$ is the flow speed, and we consider $f$ a free parameter in this model, with $f^{2}$ being the fraction of the kinetic energy over to magnetic energy ($f=1$ represents the equipartition case). Though we consider $f \le 1$ as an physically reasonable representation of the field strengths obtainable by local dynamo action (e.g. \cite{Rempel:2014ApJ...789..132R}), we also investigate some larger values to mimic the rise of magnetic field from lower depths of the convection zone to inject field into the sub-photospheric region.  

\begin{figure}[t]
\centering
\includegraphics[width=\columnwidth]{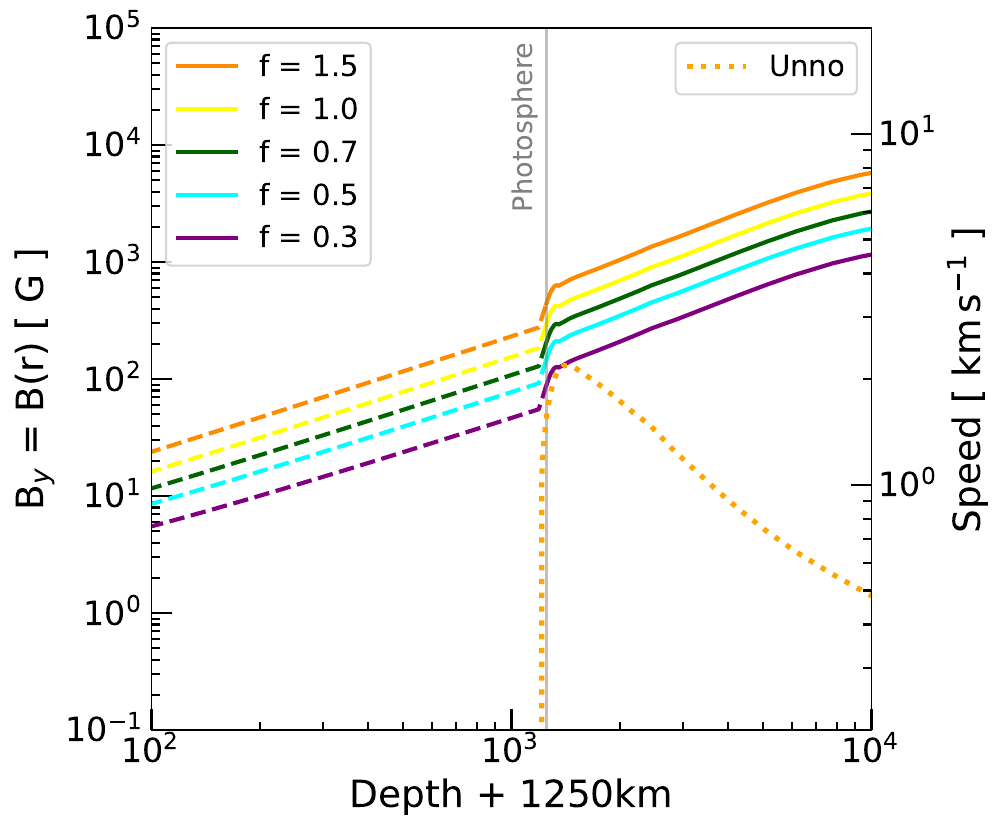}
\caption{(Left axis) Magnetic field models used in this calculation. The photosphere is located at a depth of 1250\,km. Above the photosphere, the \oneG\, case extrapolates the field to 1\,G, while the \zeroG\, case sets the field to be zero. (Right axis) The flow speed model used in this work from Ref.~\cite{Unno:1985PASJ}.  }
\label{fig:Bmodel}
\end{figure}

The speed amplitude can be estimated by convection models~\cite{Unno:1985PASJ, Deng_2006ApJ...643..426D, Xiong:2021FrASS...7...95X}, inferred from simulations~\cite{Stein_1998ApJ...499..914S}, or indirectly probed through surface observation and helioseismology~\cite{Hotta_2023SSRv..219...77H}.  
For our toy magnetic field model, we obtain the speed amplitude as a function of depth from Unno, Kondo, and Xiong~\cite{Unno:1985PASJ}.  Fig.~\ref{fig:Bmodel} right axis shows the velocity amplitude from this model, which is consistent with that from global and local simulations~(see Fig.~4 of Ref.~\cite{Miesch:2012ApJ...757..128M}).  

%%%%%%%%%%%%%%%%%%%%%%%%%%%%%%%%%%%%%%%% PLOTS
%%%%%%%%%%%%%%%%%%%%%%%%%%%%%%%%%%%%%%%% PLOTS

Figure~\ref{fig:Bmodel} shows the magnitude of the magnetic fields we used in our horizontal field model, following Eq.~(\ref{eq:bmodel}).
Above the photosphere, the magnetic fields get weaker and become more complicated. In the spirit of this paper, we consider two cases to estimate the uncertainties associated with magnetic fields above the photosphere. One is that we extrapolate the field strength linearly to 1\,G at 1250\,km above the photosphere while keeping the field direction (labeled as \oneG, dashed lines in Fig.~\ref{fig:Bmodel}); the second case is that we ignore the fields completely above the photosphere (labeled as \zeroG). These two models roughly bracket the two extreme effects of magnetic fields above the photosphere: one suppresses cosmic rays from reaching the photosphere and one have no effect on that.

\subsection{Cosmic ray orbit simulation}

Once the magnetic field model is specified, we can solve for trajectories of particles that go into the Sun, which is necessary for calculating the interaction probabilities of these particles and hence the gamma-ray production efficiencies. 
We solve the particle trajectories using the Lorentz force with the following assumptions: (1) The horizontal fields are coherent within the convection cell, and the cell is always large enough to contain our cosmic-ray particles;  (2) particle trajectories are not affected by other factors, such as interactions;  and (3) the cosmic rays are made of protons only. (The effect of nuclei on the flux is estimated separately below.)
%%%%%%%%%%%%%%%%%%%%%%%%%%%%%%%%%%%%%%%% PLOTS
\begin{figure}[h]
\centering
\includegraphics[width=\columnwidth]{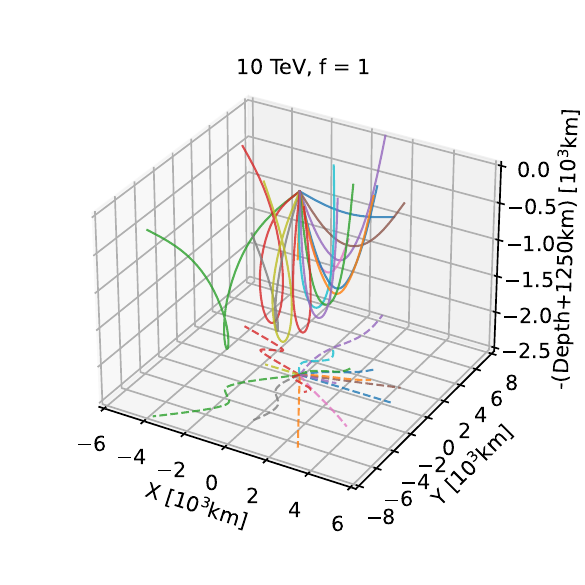}
\caption{An example of the orbit simulation, showing 15 of the 368 tracks used for the calculation for the \oneG\, case with $f = 1$ and 10\,TeV proton energy. The dashed lines are the tracks projected to the $z = -2.5\times 10^{3}$\,km plane. }
\label{fig:orbit}
\end{figure}
%%%%%%%%%%%%%%%%%%%%%%%%%%%%%%%%%%%%%%%% PLOTS

The protons are injected at 1250\,km above the photosphere, and is assumed to be isotropic. This is in part justified as Refs.~\cite{Mazziotta:2020uey, Li:2020gch} show that corona fields have a small effect on the gamma-ray production at TeV. At the injection point, a set of downgoing trajectories are simulated, assuming equal probability in pointing directions.  In practice, only the hemisphere towards the Sun is needed, as the other half will simply escape. 
The anisotropy induced by the corona field may affect the gamma-ray production when coupled to the sub-photospheric fields, however. We defer this to future studies.

Figure~\ref{fig:orbit} shows the particle trajectories obtained following the above assumptions, and by solving the equation of motion: 
\begin{equation}
    \frac{d\vec{p}}{dt} = q (\vec{\beta} \times \vec{B}) \, ,
\end{equation}
where $\vec{p}$ is the particle momentum, $\vec{\beta}$ is the particle velocity in unit of speed of light, and $q$ is the particle charge. 
The \oneG\, model is used with $f = 1$ and the particles have an energy of 10\,TeV.
A total of 368 down-going trajectories are solved for each energies, prescribed using the \texttt{healpy} package with \texttt{Nside = 8}. 
This corresponds to 368 equal-solid angle sky areas, each covers a sky area of about $(7.3\,{\rm deg})^{2}$.
For each trajectory, labeled by the solid angle $\Omega$, we obtain its total optical depth along the track, $\tau(E_{p}, \Omega) = \int \frac{\rho}{m_{p}} \sigma d\ell$, where $\ell$ is the track length element.

\begin{figure}[t]
\centering
\includegraphics[width=\columnwidth]{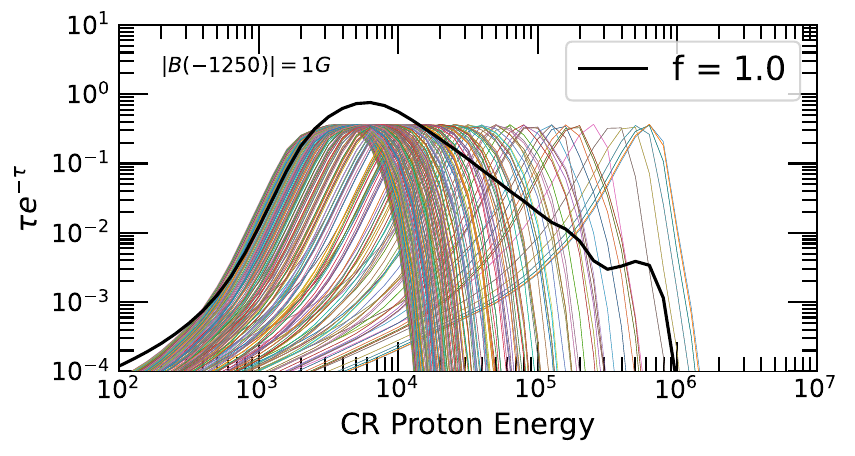}
\includegraphics[width=\columnwidth]{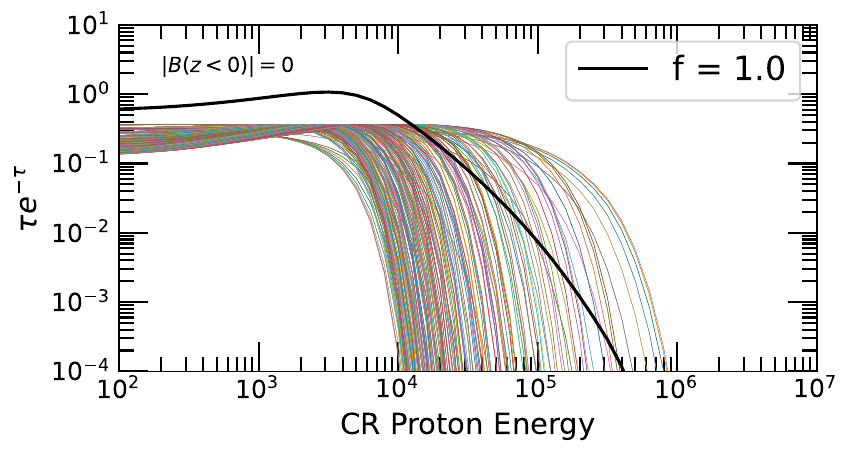}
\caption{Colored lines show the efficiency factor ${\cal T} = \tau e^{-\tau}$ as a function of cosmic-ray proton energies for different downgoing trajectories.  The black solid line is the zenith-weighted sum of the efficiencies $\int \cos\theta {\cal T} d\Omega$. The top panel represents the \oneG\, case, and the bottom panel corresponds to the \zeroG\, case.  }
\label{fig:efficiencies}
\end{figure}

Figure~\ref{fig:efficiencies} shows gamma-ray efficiency factor for the tracks of the \oneG\, and \zeroG\, cases, respectively. 
The efficiency factor is defined as by 
\begin{equation}\label{eq:opticaldepth}
{\cal T}(E_{p}, \Omega) = \tau(E_{p}, \Omega) e^{-\tau(E_{p}, \Omega)} \, 
\end{equation}
using the optical depth solutions obtained from particle trajectories above. 
Given the field geometry, the particles will always be reflected and hence contribute to the gamma ray production.  We leave subtle cases, such as particles interacting before reflection, etc, for future numerical follow ups.  
The exponential factor in Eq.~\ref{eq:opticaldepth} approximates the effectiveness of particle attenuation in high-optical depth scenarios.  

The solid lines in Fig.~\ref{fig:efficiencies} are the zenith-weighted sum $\int \cos\theta \tau e^{-\tau} d\Omega$. Here we can see the main differences between the two magnetic field models.  For the \oneG\, case, the efficiency is suppressed at low energies as cosmic rays can be reflected off the Sun before reaching sufficient column density to interact. This suppression is especially significant for particles with small inclination angles.  

The efficiency also drops off at high energies due to the $e^{-\tau}$ factor. The efficiency peak ($1/e$) is located at higher energies for the less inclined trajectories as they transverse the shallower part of the atmosphere. 
For the \zeroG\, case, as the particles only experience the magnetic fields after they enter the photosphere, even low energy particles could accumulate sufficient column density in the high density atmosphere. And similarly, less inclined particles contributes to the efficiency at higher energies.

%%%%%%%%%%%%%%%%%%%%%%%%%%%%%%%%%%%%%%%% PLOTS
%%%%%%%%%%%%%%%%%%%%%%%%%%%%%%%%%%%%%%%% PLOTS

% % ====== Formulation ====== %
\section{\label{sec:formulation} Gamma-ray flux production}

Given the orbit information, we can estimate the gamma-ray production semi-analytically. 
We first write down the emissivity following the delta function approximation from Ref.~\cite{Kelner:2006tc}. The emissivity here is the number of photons produced per energy per surface area per incoming solid angle,  
\begin{eqnarray}
\frac{dN}{dE_{\gamma} d\Omega dA}(E_{p}, \Omega) &=& 
N_{\rm nuc} \frac{2c }{K_{\pi}} \int_{ E_{\rm min}}  dE_{\pi} {\cal T}(\tilde{E}, \Omega)   
\nonumber\\&&\times
\frac{\Phi_{p}( \tilde{E}) }{\sqrt{E_{\pi}^2 - m_{\pi}^2 }}\,,
\label{eq:dNdEdOdA}
\end{eqnarray}
where $N_{\rm nuc}$ is the nuclear enhancement factor, $K_{\pi} = 0.17$, $E_{\rm min} = E_{\gamma} +{m_{\pi}^{2}}/{4 E_{\gamma} }$, $\tilde{E} = m_{p} + {E_{\pi}}/{K_{\pi}}$, and $\Phi_{p}(E_{p})$ is the cosmic-ray proton flux, for which we use the analytic double break expression from Ref.~\cite{Lipari:2019jmk}, which is a fit of cosmic-ray data from 0.1\,TeV to 100\,TeV at Earth position. At this energy, we expect the effect of cosmic-ray modulation to be negligible~\cite{Li:2022zio}.  
The nuclear enhancement factor $N_{\rm nuc}$ takes into account the enhanced gamma-ray production due to the presence of nuclei in the cosmic rays and the target~\cite{Mori:2009te, Kachelriess:2014mga}. The only relevant species to be considered is Helium, which is $< 10$\% in both cases.  We follow Ref.~\cite{Zhou:2016ljf} and take the enhancement factor as a constant of 1.8.

The total photon flux is then obtained by integrating Eq.~(\ref{eq:dNdEdOdA}) over the solid angle and the surface area, $4\pi R^{2}$, divided by $4\pi D^{2}$, hence,
\begin{equation}
\frac{dF}{dE} =  F_{\rm sur} \frac{R^{2}}{D^2} \int d\Omega \cos\theta \frac{dN}{dE_{\gamma} d\Omega dA}(E_{p}, \Omega)\, ,
\label{eq:dFdE}
\end{equation}
where $F_{\rm sur}$ is the surface fraction that the cosmic rays would encounter the internetwork fields, and $\cos\theta$ takes the projection effect between the surface area element and the cosmic-ray incoming angle into account. Here we have taken the cosmic-ray flux to be isotropic at 1250\,km above the photosphere. 

\begin{figure}[t]
\centering
\includegraphics[width=\columnwidth]{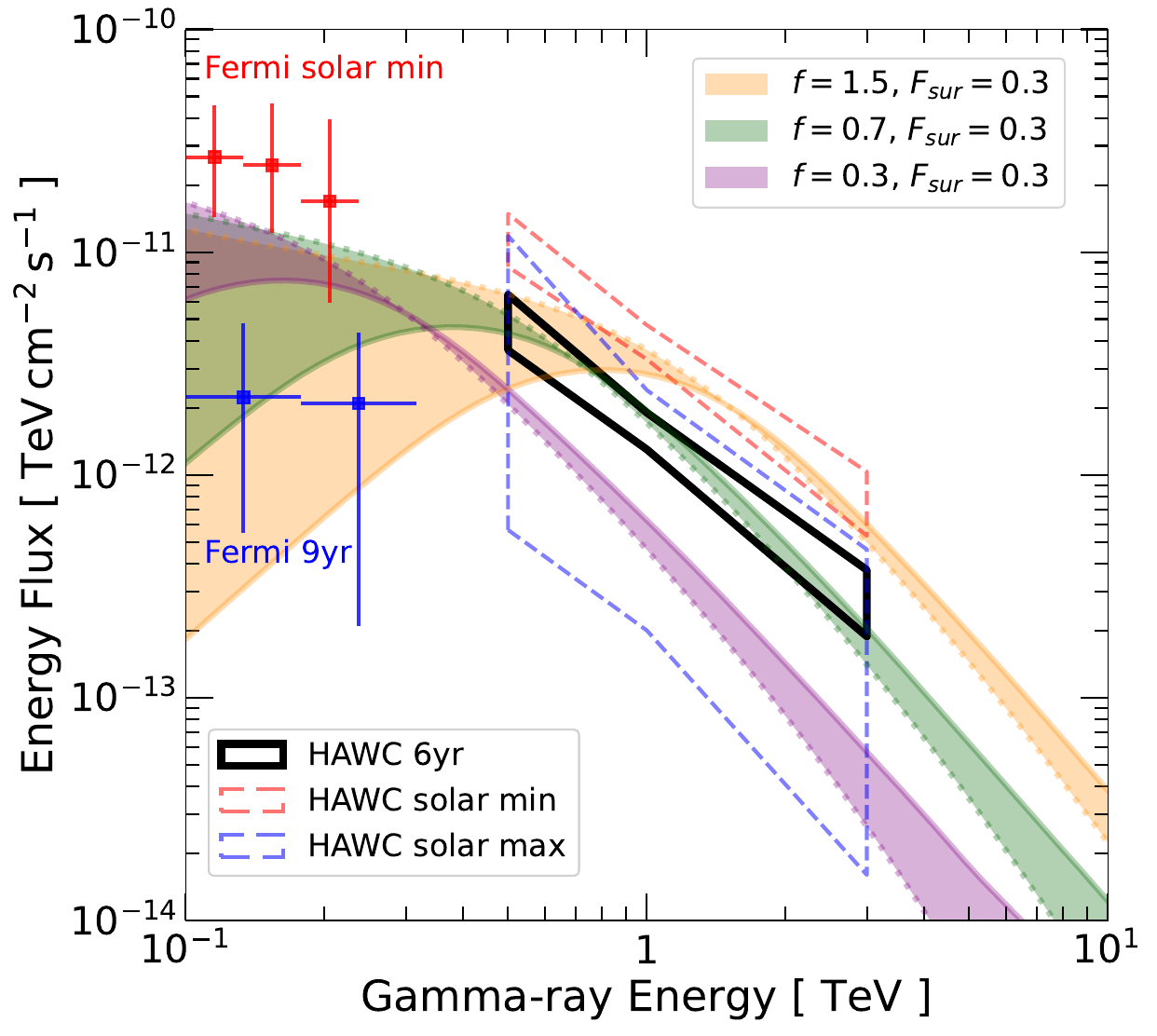}
\caption{Gamma ray flux produced by cosmic rays reflecting on the horizontal internetwork fields, for $f =  0.3$, 0.7, and 1.5.  The flux is directly proportional to the surface fraction $F_{surf}$ [Eq.~(\ref{eq:dFdE})], which is taken to be $1$ in this figure.  
Dotted lines and solid lines correspond to the \zeroG and \oneG cases, respectively. 
The Fermi-LAT data at the solar minimum (2008-2010) and 9-year (2008-2017) averaged flux data~\cite{Tang:2018wqp} are shown for comparison.  
The recent HAWC results~\cite{HAWC:2022khj} are shown in thick boxes, with black solid line being the 6-year~(2014-2021) combined result, blue dashed line correspond to solar max. (2014-2017), and red dashed line correspond to the solar min. (2018-2021) It can be seen that our model could reasonably explain the HAWC data, e.g., with $f \simeq 0.7$ and $F_{sur} \simeq 0.3$. }
\label{fig:results}
\end{figure}

Figure~\ref{fig:results} shows the computed solar gamma-ray spectra for $f = 0.3$, 0.7, and 1.5, and $F_{\rm sur} = 0.3$.  
Considering the \oneG\, case, at lower energies, the gamma-ray production are suppressed compared to the \zeroG\, case. 
At higher energies, the additional magnetic fields provides extra deflections that shift the efficiency factor towards higher energies, thus making the \oneG\, case higher. 

From Fig.~\ref{fig:results}, we can see that horizontal internetwork magnetic fields could explain the HAWC 6-yr observations with a reasonable choice of parameters $f \simeq 0.7$ and $F_{\rm sur} \simeq 0.3$. We note that these two parameters are somewhat degenerate if only considering the results above 1\,TeV.

%%%%%%%%%%%%%%%%%%%%%%%%%%%%%%%%%%%%%%%% PLOTS
%%%%%%%%%%%%%%%%%%%%%%%%%%%%%%%%%%%%%%%% 

% % ====== Results and Discussion ====== %
\section{\label{sec:results} Conclusion and Discussion}

In this work, we provide the first physical model that can fully explain the HAWC TeV solar observation.  We show that horizontal sub-photospheric internetwork fields, an extension of those seen at the photosphere, could be strong and ubiquitous enough to reflect $\sim$TeV cosmic rays and generate the observed solar gamma rays.   
Our model is simple and has few parameters, as the magnetic field strength can be reasonably connected to the kinetic energy of the fluid through the scaling parameter $f$.  Another parameter is the fraction of the Sun that are covered by the horizontal fields, $F_{\rm sur}$. 
A reasonable choice of parameters $f \simeq 0.7$ and $F_{\rm sur} \simeq 0.3$ could explain the HAWC 6-yr observations, for instance. 

Nevertheless, our model has some shortcomings that likely can only be addressed with detailed numerical investigations with more realistic setup.  For instance, we assume that the particles traverse the same convection cell with the coherent field direction, which is probably a good assumption for nearly vertical trajectories, but less so for the ones with small incident angles that could travel across cells.  Particles traveling across cells may encounter vertical fields in cell boundaries and different horizontal field orientations in other cells, which would likely prolong the time the particle stays in the photosphere, and could soften the spectrum at high energies. A more realistic convection cell model with different sizes and magnetic field orientations would be needed to address this. In addition, the details of particle shower development in the Sun under magnetic fields can only be tackled in a numerical setup, which could affect the gamma-ray production yield at lower energies substantially.  

We focus on the TeV energy range, which allows us to ignore effects from weaker magnetic fields.  To fully explain the solar gamma-ray observation from TeV down to the GeV energy range, a comprehensive model that includes solar modulations, corona magnetic fields, and network magnetic fields is likely necessary. Notably, Sun shadow observations might be used to isolate the effect of large scale magnetic fields (e.g., from the corona)~\cite{Amenomori:2013own, Tibet:2018PhRvL.120c1101A, sunshadowAartsen:2018zrl, sunshadowBeckerTjus:2019rqu, sunshadowAartsen:2020hzn}. 

While our model could explain the TeV flux data, a slew of observational features still need to be addressed, such as time-dependence~\cite{Ng:2015gya, Tang:2018wqp, Linden:2020lvz, HAWC:2022khj} and surface morphology at the Fermi energy range~\cite{Linden:2018exo, Arsioli:2024scu}.   
For these observations, one would require a better understanding on how the related magnetic fields changes in time and surface distribution. The morphology also depends on the angular distribution of the emitted photons after cosmic-ray reflection and interaction, which can only be tracked with numerical simulation. 

The understanding of TeV solar gamma rays is also important for predicting solar atmospheric neutrino flux~\cite{Ingelman:1996mj, Ng:2017aur, Arguelles:2017eao, Edsjo:2017kjk}, which could be detectable in neutrino telescopes~\cite{Aartsen:2019avh_solaratmneutrino, IceCube:2021koo_icrcsolar} and has strong implications for new physics searches~(e.g., see Ref.~\cite{Aartsen:2016zhm_darkmatter, Leane:2017vag, Niblaeus:2019gjk}).

In summary, we show that TeV gamma rays could be a novel probe for the magnetic fields and the fluid dynamics just under the photosphere, which are difficult to probe with traditional observational methods.  This will be enabled by continued monitoring of the Sun by HAWC, LHAASO, and SWGO~\cite{SWGO:2019ahw}, as well as closer theory-data connection with future numerical work on the solar gamma-ray production.

% ====== ACKNOWLEDGEMENTS ====== %
\section*{\label{sec:acknowledgements} Acknowledgments}
 KCYN is supported by Croucher foundation, RGC grants (24302721, 14305822, 14308023), Joint NSFC/GRC grant (N CUHK456/22), and NSFC grant 12322517. 
AH is supported by STFC Research Grant No. ST/V000659/1.
The work of SA was supported by JSPS/MEXT KAKENHI under grant numbers JP20H05850, JP20H05861, and JP24K07039.
AH would like to acknowledge conversations with Profs Oskar Steiner and Hideyuki Hotta which helped with the generation of the magnetic field model. 
KCYN would like to thank John Beacom, Hugh Hudson, Federico Fraschetti, Ofer Cohen, Eleonora Puzzoni, Annika Peter, and Jung-Tsung Li for helpful comments and discussions.
For the purpose of open access, the author has applied a ‘Creative Commons Attribution (CC BY) licence to any Author Accepted Manuscript version arising.

% ====== bib ====== %
\bibliography{bib.bib}

\providecommand{\href}[2]{#2}\begingroup\raggedright\begin{thebibliography}{10}

\bibitem{Seckel:1991ffa}
D.~Seckel, T.~Stanev and T.~K. Gaisser, \emph{{Signatures of cosmic-ray
  interactions on the solar surface}},
  \href{http://dx.doi.org/10.1086/170753}{\emph{Astrophys. J.} {\bf 382} (1991)
  652--666}.

\bibitem{Zhou:2016ljf}
B.~Zhou, K.~C.~Y. Ng, J.~F. Beacom and A.~H.~G. Peter, \emph{{TeV Solar Gamma
  Rays From Cosmic-Ray Interactions}},
  \href{http://dx.doi.org/10.1103/PhysRevD.96.023015}{\emph{Phys. Rev.} {\bf
  D96} (2017) 023015}, [\href{http://arxiv.org/abs/1612.02420}{{\tt
  1612.02420}}].

\bibitem{Orlando:2008uk}
E.~Orlando and A.~W. Strong, \emph{{Gamma-ray emission from the solar halo and
  disk: a study with EGRET data}},
  \href{http://dx.doi.org/10.1051/0004-6361:20078817}{\emph{Astron. Astrophys.}
  {\bf 480} (2008) 847}, [\href{http://arxiv.org/abs/0801.2178}{{\tt
  0801.2178}}].

\bibitem{Abdo:2011xn}
{\scshape Fermi-LAT} collaboration, A.~A. Abdo et~al., \emph{{Fermi-LAT
  Observations of Two Gamma-Ray Emission Components from the Quiescent Sun}},
  \href{http://dx.doi.org/10.1088/0004-637X/734/2/116}{\emph{Astrophys. J.}
  {\bf 734} (2011) 116}, [\href{http://arxiv.org/abs/1104.2093}{{\tt
  1104.2093}}].

\bibitem{Ng:2015gya}
K.~C.~Y. Ng, J.~F. Beacom, A.~H.~G. Peter and C.~Rott, \emph{{First Observation
  of Time Variation in the Solar-Disk Gamma-Ray Flux with Fermi}},
  \href{http://dx.doi.org/10.1103/PhysRevD.94.023004}{\emph{Phys. Rev.} {\bf
  D94} (2016) 023004}, [\href{http://arxiv.org/abs/1508.06276}{{\tt
  1508.06276}}].

\bibitem{Tang:2018wqp}
Q.-W. Tang, K.~C.~Y. Ng, T.~Linden, B.~Zhou, J.~F. Beacom and A.~H.~G. Peter,
  \emph{{Unexpected dip in the solar gamma-ray spectrum}},
  \href{http://dx.doi.org/10.1103/PhysRevD.98.063019}{\emph{Phys. Rev.} {\bf
  D98} (2018) 063019}, [\href{http://arxiv.org/abs/1804.06846}{{\tt
  1804.06846}}].

\bibitem{Linden:2018exo}
T.~Linden, B.~Zhou, J.~F. Beacom, A.~H.~G. Peter, K.~C.~Y. Ng and Q.-W. Tang,
  \emph{{Evidence for a New Component of High-Energy Solar Gamma-Ray
  Production}},
  \href{http://dx.doi.org/10.1103/PhysRevLett.121.131103}{\emph{Phys. Rev.
  Lett.} {\bf 121} (2018) 131103}, [\href{http://arxiv.org/abs/1803.05436}{{\tt
  1803.05436}}].

\bibitem{Linden:2020lvz}
T.~Linden, J.~F. Beacom, A.~H.~G. Peter, B.~J. Buckman, B.~Zhou and G.~Zhu,
  \emph{{First observations of solar disk gamma rays over a full solar cycle}},
  \href{http://dx.doi.org/10.1103/PhysRevD.105.063013}{\emph{Phys. Rev. D} {\bf
  105} (2022) 063013}, [\href{http://arxiv.org/abs/2012.04654}{{\tt
  2012.04654}}].

\bibitem{Arsioli:2024scu}
B.~Arsioli and E.~Orlando, \emph{{Yet Another Sunshine Mystery: Unexpected
  Asymmetry in GeV Emission from the Solar Disk}},
  \href{http://dx.doi.org/10.3847/1538-4357/ad1bd2}{\emph{Astrophys. J.} {\bf
  962} (2024) 52}, [\href{http://arxiv.org/abs/2401.03466}{{\tt 2401.03466}}].

\bibitem{HAWC:2022khj}
{\scshape HAWC} collaboration, A.~Albert et~al., \emph{{Discovery of Gamma Rays
  from the Quiescent Sun with HAWC}},
  \href{http://dx.doi.org/10.1103/PhysRevLett.131.051201}{\emph{Phys. Rev.
  Lett.} {\bf 131} (2023) 051201}, [\href{http://arxiv.org/abs/2212.00815}{{\tt
  2212.00815}}].

\bibitem{Mazziotta:2020uey}
M.~N. Mazziotta, P.~De~La Torre~Luque, L.~Di~Venere, A.~Fass\`o, A.~Ferrari,
  F.~Loparco et~al., \emph{{Cosmic-ray interactions with the Sun using the
  FLUKA code}},
  \href{http://dx.doi.org/10.1103/PhysRevD.101.083011}{\emph{Phys. Rev. D} {\bf
  101} (2020) 083011}, [\href{http://arxiv.org/abs/2001.09933}{{\tt
  2001.09933}}].

\bibitem{Li:2020gch}
Z.~Li, K.~C.~Y. Ng, S.~Chen, Y.~Nan and H.~He, \emph{{Simulating gamma-ray
  production from cosmic rays interacting with the solar atmosphere in the
  presence of coronal magnetic fields*}},
  \href{http://dx.doi.org/10.1088/1674-1137/ad1cda}{\emph{Chin. Phys. C} {\bf
  48} (2024) 045101}, [\href{http://arxiv.org/abs/2009.03888}{{\tt
  2009.03888}}].

\bibitem{Schatten:1969}
K.~H. {Schatten}, J.~M. {Wilcox} and N.~F. {Ness}, \emph{{A model of
  interplanetary and coronal magnetic fields}},
  \href{http://dx.doi.org/10.1007/BF00146478}{\emph{Sol Phys} {\bf 6} (1969)
  442}.

\bibitem{Altschuler:1969SoPh}
M.~D. {Altschuler} and G.~{Newkirk}, \emph{{Magnetic Fields and the Structure
  of the Solar Corona. I: Methods of Calculating Coronal Fields}},
  \href{http://dx.doi.org/10.1007/BF00145734}{\emph{Sol.~Phys.} {\bf 9} (Sept.,
  1969) 131--149}.

\bibitem{Hoeksema:1984PhDT}
J.~T. {Hoeksema}, \emph{{Structure and Evolution of the Large Scale Solar and
  Heliospheric Magnetic Fields.}}
\newblock PhD thesis, Stanford Univ., CA., Sept., 1984.

\bibitem{Wang:1992ApJ}
Y.~M. {Wang} and J.~{Sheeley}, N.~R., \emph{{On Potential Field Models of the
  Solar Corona}}, \href{http://dx.doi.org/10.1086/171430}{\emph{\apj} {\bf 392}
  (June, 1992) 310}.

\bibitem{Gutierrez:2019fna}
M.~Guti{\'e}rrez and M.~Masip, \emph{{The Sun at TeV energies: gammas,
  neutrons, neutrinos and a cosmic ray shadow}},
  \href{http://dx.doi.org/10.1016/j.astropartphys.2020.102440}{\emph{Astropart.
  Phys.} {\bf 119} (2020) 102440}, [\href{http://arxiv.org/abs/1911.07530}{{\tt
  1911.07530}}].

\bibitem{Gutierrez:2022mor}
M.~Guti\'errez, M.~Masip and S.~Mu\~noz, \emph{{The Solar Disk at High
  Energies}},
  \href{http://dx.doi.org/10.3847/1538-4357/aca020}{\emph{Astrophys. J.} {\bf
  941} (2022) 86}, [\href{http://arxiv.org/abs/2206.00964}{{\tt 2206.00964}}].

\bibitem{Banik:2023shc}
P.~Banik, A.~Bhadra and S.~K. Ghosh, \emph{{Sun as a cosmic ray TeVatron}},
  \href{http://arxiv.org/abs/2305.17086}{{\tt 2305.17086}}.

\bibitem{Hudson2020MNRAS.491.4852H}
H.~S. {Hudson}, A.~{MacKinnon}, M.~{Szydlarski} and M.~{Carlsson},
  \emph{{Cosmic ray interactions in the solar atmosphere}},
  \href{http://dx.doi.org/10.1093/mnras/stz3373}{\emph{MNRAS} {\bf 491} (Feb.,
  2020) 4852--4856}, [\href{http://arxiv.org/abs/1910.01186}{{\tt
  1910.01186}}].

\bibitem{Li:2023twp}
J.-T. Li, J.~F. Beacom, S.~Griffith and A.~H.~G. Peter, \emph{{Small-scale
  Magnetic Fields Are Critical to Shaping Solar Gamma-Ray Emission}},
  \href{http://dx.doi.org/10.3847/1538-4357/ad158f}{\emph{Astrophys. J.} {\bf
  961} (2024) 167}, [\href{http://arxiv.org/abs/2307.08728}{{\tt 2307.08728}}].

\bibitem{Ramaty1975SSRv...18..341R}
R.~{Ramaty}, B.~{Kozlovsky} and R.~E. {Lingenfelter}, \emph{{Solar Gamma
  Rays}}, \href{http://dx.doi.org/10.1007/BF00212911}{\emph{Space~Sci.~Rev.}
  {\bf 18} (Dec., 1975) 341--388}.

\bibitem{Murphy1987ApJS...63..721M}
R.~J. {Murphy}, C.~D. {Dermer} and R.~{Ramaty}, \emph{{High-Energy Processes in
  Solar Flares}}, \href{http://dx.doi.org/10.1086/191180}{\emph{ApJS} {\bf 63}
  (Mar., 1987) 721}.

\bibitem{Kozlovsky2002ApJS..141..523K}
B.~{Kozlovsky}, R.~J. {Murphy} and R.~{Ramaty}, \emph{{Nuclear Deexcitation
  Gamma-Ray Lines from Accelerated Particle Interactions}},
  \href{http://dx.doi.org/10.1086/340545}{\emph{ApJS} {\bf 141} (Aug., 2002)
  523--541}.

\bibitem{Omodei:2018uni}
N.~Omodei, M.~Pesce-Rollins, F.~Longo, A.~Allafort and S.~Krucker,
  \emph{{Fermi-LAT Observations of the 2017 September 10 Solar Flare}},
  \href{http://dx.doi.org/10.3847/2041-8213/aae077}{\emph{Astrophys. J. Lett.}
  {\bf 865} (2018) L7}, [\href{http://arxiv.org/abs/1803.07654}{{\tt
  1803.07654}}].

\bibitem{Share:2018kqt}
G.~H. Share, R.~J. Murphy, S.~M. White, A.~K. Tolbert, B.~R. Dennis, R.~A.
  Schwartz et~al., \emph{{Characteristics of Late-phase \ensuremath{>}100 MeV
  Gamma-Ray Emission in Solar Eruptive Events}},
  \href{http://dx.doi.org/10.3847/1538-4357/aaebf7}{\emph{Astrophys. J.} {\bf
  869} (2018) 182}.

\bibitem{Fermi-LAT:2013vao_flares}
{\scshape Fermi-LAT} collaboration, M.~Ackermann et~al., \emph{{High-energy
  Gamma-Ray Emission from Solar Flares: Summary of Fermi Large Area Telescope
  Detections and Analysis of Two M-class Flares}},
  \href{http://dx.doi.org/10.1088/0004-637X/787/1/15}{\emph{Astrophys. J.} {\bf
  787} (2014) 15}, [\href{http://arxiv.org/abs/1304.3749}{{\tt 1304.3749}}].

\bibitem{Orlando:2006zs}
E.~Orlando and A.~Strong, \emph{{Gamma-rays from halos around stars and the
  Sun}}, \href{http://dx.doi.org/10.1007/s10509-007-9457-0}{\emph{Astrophys.
  Space Sci.} {\bf 309} (2007) 359--363},
  [\href{http://arxiv.org/abs/astro-ph/0607563}{{\tt astro-ph/0607563}}].

\bibitem{Moskalenko:2006ta}
I.~V. Moskalenko, T.~A. Porter and S.~W. Digel, \emph{{Inverse Compton
  scattering on solar photons, heliospheric modulation, and neutrino
  astrophysics}}, \href{http://dx.doi.org/10.1086/520882,
  10.1086/509916}{\emph{Astrophys. J.} {\bf 652} (2006) L65--L68},
  [\href{http://arxiv.org/abs/astro-ph/0607521}{{\tt astro-ph/0607521}}].

\bibitem{Orlando:2020ezh}
E.~Orlando and A.~Strong, \emph{{StellarICS: Inverse Compton Emission from the
  Quiet Sun and Stars from keV to TeV}},
  \href{http://dx.doi.org/10.1088/1475-7516/2021/04/004}{\emph{JCAP} {\bf 04}
  (2021) 004}, [\href{http://arxiv.org/abs/2012.13126}{{\tt 2012.13126}}].

\bibitem{Lai:2022qif}
A.~C.~M. Lai and K.~C.~Y. Ng, \emph{{Anisotropic photon and electron scattering
  without ultrarelativistic approximation}},
  \href{http://dx.doi.org/10.1103/PhysRevD.107.063026}{\emph{Phys. Rev. D} {\bf
  107} (2023) 063026}, [\href{http://arxiv.org/abs/2211.15691}{{\tt
  2211.15691}}].

\bibitem{Yang:2023res}
H.-G. Yang, Y.~Gao, Y.-Z. Ma and R.~M. Crocker, \emph{{Solar gamma ray probe of
  local cosmic ray electrons}},
  \href{http://dx.doi.org/10.1103/PhysRevD.108.L061304}{\emph{Phys. Rev. D}
  {\bf 108} (2023) L061304}, [\href{http://arxiv.org/abs/2309.04784}{{\tt
  2309.04784}}].

\bibitem{Petrosian:2022bxr}
V.~Petrosian, E.~Orlando and A.~Strong, \emph{{Transport of Cosmic-Ray
  Electrons from 1 au to the Sun}},
  \href{http://dx.doi.org/10.3847/1538-4357/aca474}{\emph{Astrophys. J.} {\bf
  943} (2023) 21}, [\href{http://arxiv.org/abs/2212.00929}{{\tt 2212.00929}}].

\bibitem{Orlando:2022xsm}
E.~Orlando, V.~Petrosian and A.~Strong, \emph{{A New Component from the Quiet
  Sun from Radio to Gamma Rays: Synchrotron Radiation by Galactic Cosmic-Ray
  Electrons}},
  \href{http://dx.doi.org/10.3847/1538-4357/acad75}{\emph{Astrophys. J.} {\bf
  943} (2023) 173}, [\href{http://arxiv.org/abs/2212.01364}{{\tt 2212.01364}}].

\bibitem{Baker:1966}
{Norman H. Baker, Stefan Temesv{\'a}ry}, \emph{{Tables of Convective Stellar
  Envelope Models}}.
\newblock 1966.

\bibitem{1973ApJ...184..605V}
J.~E. {Vernazza}, E.~H. {Avrett} and R.~{Loeser}, \emph{{Structure of the Solar
  Chromosphere. Basic Computations and Summary of the Results}},
  \href{http://dx.doi.org/10.1086/152353}{\emph{\apj} {\bf 184} (Sept., 1973)
  605--632}.

\bibitem{Kelner:2006tc}
S.~R. Kelner, F.~A. Aharonian and V.~V. Bugayov, \emph{{Energy spectra of
  gamma-rays, electrons and neutrinos produced at proton-proton interactions in
  the very high energy regime}},
  \href{http://dx.doi.org/10.1103/PhysRevD.74.034018}{\emph{Phys. Rev. D} {\bf
  74} (2006) 034018}, [\href{http://arxiv.org/abs/astro-ph/0606058}{{\tt
  astro-ph/0606058}}].

\bibitem{Workman:2022ynf}
{\scshape Particle Data Group} collaboration, R.~L. Workman and Others,
  \emph{{Review of Particle Physics}},
  \href{http://dx.doi.org/10.1093/ptep/ptac097}{\emph{PTEP} {\bf 2022} (2022)
  083C01}.

\bibitem{Unno:1985PASJ}
W.~{Unno}, M.~A. {Kondo} and D.~R. {Xiong}, \emph{{Solar convection zone given
  by nonlocal mixing-length theory}}, {\emph{\rm PASJ} {\bf 37} (Jan., 1985)
  235--244}.

\bibitem{Wedemeyer-Bohm:2008tpk}
S.~Wedemeyer-Bohm, A.~Lagg and A.~Nordlund, \emph{{Coupling from the
  photosphere to the chromosphere and the corona}},
  \href{http://dx.doi.org/10.1007/s11214-008-9447-8}{\emph{Space Sci. Rev.}
  {\bf 144} (2009) 317--350}, [\href{http://arxiv.org/abs/0809.0987}{{\tt
  0809.0987}}].

\bibitem{Wiegelmann2014}
T.~{Wiegelmann}, J.~K. {Thalmann} and S.~K. {Solanki}, \emph{{The magnetic
  field in the solar atmosphere}},
  \href{http://dx.doi.org/10.1007/s00159-014-0078-7}{\emph{Astronomy and
  Astrophysics Reviews} {\bf 22} (Nov., 2014) 78},
  [\href{http://arxiv.org/abs/1410.4214}{{\tt 1410.4214}}].

\bibitem{Solanki2006RPPh...69..563S}
S.~K. {Solanki}, B.~{Inhester} and M.~{Sch{\"u}ssler}, \emph{{The solar
  magnetic field}},
  \href{http://dx.doi.org/10.1088/0034-4885/69/3/R02}{\emph{Reports on Progress
  in Physics} {\bf 69} (Mar., 2006) 563--668},
  [\href{http://arxiv.org/abs/1008.0771}{{\tt 1008.0771}}].

\bibitem{Almeida:2011talk}
J.~{S{\'a}nchez Almeida} and M.~{Mart{\'\i}nez Gonz{\'a}lez}, \emph{{The
  Magnetic Fields of the Quiet Sun}},  in \emph{Solar Polarization 6} (J.~R.
  {Kuhn}, D.~M. {Harrington}, H.~{Lin}, S.~V. {Berdyugina},
  J.~{Trujillo-Bueno}, S.~L. {Keil} et~al., eds.), vol.~437 of
  \emph{Astronomical Society of the Pacific Conference Series}, p.~451, Apr.,
  2011.
\newblock \href{http://arxiv.org/abs/1105.0387}{{\tt 1105.0387}}.
\newblock \href{http://dx.doi.org/10.48550/arXiv.1105.0387}{DOI}.

\bibitem{TrujilloBueno:2004re}
J.~Trujillo~Bueno, N.~Shchukina and A.~Asensio~Ramos, \emph{{A Substantial
  amount of hidden magnetic energy in the quiet Sun}},
  \href{http://dx.doi.org/10.1038/nature02669}{\emph{Nature} {\bf 430} (2004)
  326--329}, [\href{http://arxiv.org/abs/astro-ph/0409004}{{\tt
  astro-ph/0409004}}].

\bibitem{Rubio2019LRSP}
L.~{Bellot Rubio} and D.~{Orozco Su{\'a}rez}, \emph{{Quiet Sun magnetic fields:
  an observational view}},
  \href{http://dx.doi.org/10.1007/s41116-018-0017-1}{\emph{Living Reviews in
  Solar Physics} {\bf 16} (Feb., 2019) 1}.

\bibitem{Rempel2023SSRv}
M.~{Rempel}, T.~{Bhatia}, L.~{Bellot Rubio} and M.~J. {Korpi-Lagg},
  \emph{{Small-Scale Dynamos: From Idealized Models to Solar and Stellar
  Applications}},
  \href{http://dx.doi.org/10.1007/s11214-023-00981-z}{\emph{Space~Sci.~Rev.}
  {\bf 219} (Aug., 2023) 36}.

\bibitem{Lites1996ApJ...460.1019L}
B.~W. {Lites}, K.~D. {Leka}, A.~{Skumanich}, V.~{Martinez Pillet} and
  T.~{Shimizu}, \emph{{Small-Scale Horizontal Magnetic Fields in the Solar
  Photosphere}}, \href{http://dx.doi.org/10.1086/177028}{\emph{ApJ} {\bf 460}
  (Apr., 1996) 1019}.

\bibitem{Suarez:2007pp}
D.~O. Suarez et~al., \emph{{Quiet Sun internetwork magnetic fields from the
  inversion of Hinode measurements}},
  \href{http://dx.doi.org/10.1086/524139}{\emph{Astrophys. J. Lett.} {\bf 670}
  (2007) L61}, [\href{http://arxiv.org/abs/0710.1405}{{\tt 0710.1405}}].

\bibitem{Lites2008ApJ...672.1237L}
B.~W. {Lites}, M.~{Kubo}, H.~{Socas-Navarro}, T.~{Berger}, Z.~{Frank},
  R.~{Shine} et~al., \emph{{The Horizontal Magnetic Flux of the Quiet-Sun
  Internetwork as Observed with the Hinode Spectro-Polarimeter}},
  \href{http://dx.doi.org/10.1086/522922}{\emph{\apj} {\bf 672} (Jan., 2008)
  1237--1253}.

\bibitem{Schuessler:2008si}
M.~Schuessler and A.~Voegler, \emph{{Strong horizontal photospheric magnetic
  field in a surface dynamo simulation}},
  \href{http://dx.doi.org/10.1051/0004-6361:20078998}{\emph{Astron. Astrophys.}
  {\bf 481} (2008) L5}, [\href{http://arxiv.org/abs/0801.1250}{{\tt
  0801.1250}}].

\bibitem{Steiner2008ApJ...680L..85S}
O.~{Steiner}, R.~{Rezaei}, W.~{Schaffenberger} and S.~{Wedemeyer-B{\"o}hm},
  \emph{{The Horizontal Internetwork Magnetic Field: Numerical Simulations in
  Comparison to Observations with Hinode}},
  \href{http://dx.doi.org/10.1086/589740}{\emph{ApjL} {\bf 680} (June, 2008)
  L85}, [\href{http://arxiv.org/abs/0801.4915}{{\tt 0801.4915}}].

\bibitem{Steiner2010ASSP...19..166S}
O.~{Steiner}, \emph{{Magnetic Coupling in the Quiet Solar Atmosphere}},  in
  \emph{Magnetic Coupling between the Interior and Atmosphere of the Sun},
  vol.~19 of \emph{Astrophysics and Space Science Proceedings}, pp.~166--185,
  Jan., 2010.
\newblock \href{http://arxiv.org/abs/0904.2026}{{\tt 0904.2026}}.
\newblock \href{http://dx.doi.org/10.1007/978-3-642-02859-5_13}{DOI}.

\bibitem{Potgieter:2013pdj}
M.~Potgieter, \emph{{Solar Modulation of Cosmic Rays}},
  \href{http://dx.doi.org/10.12942/lrsp-2013-3}{\emph{Living Rev. Solar Phys.}
  {\bf 10} (2013) 3}, [\href{http://arxiv.org/abs/1306.4421}{{\tt 1306.4421}}].

\bibitem{Li:2022zio}
J.-T. Li, J.~F. Beacom and A.~H.~G. Peter, \emph{{Galactic Cosmic-Ray
  Propagation in the Inner Heliosphere: Improved Force-field Model}},
  \href{http://dx.doi.org/10.3847/1538-4357/ac8cf3}{\emph{Astrophys. J.} {\bf
  937} (2022) 27}, [\href{http://arxiv.org/abs/2206.14815}{{\tt 2206.14815}}].

\bibitem{Rempel:2014ApJ...789..132R}
M.~{Rempel}, \emph{{Numerical Simulations of Quiet Sun Magnetism: On the
  Contribution from a Small-scale Dynamo}},
  \href{http://dx.doi.org/10.1088/0004-637X/789/2/132}{\emph{\apj} {\bf 789}
  (July, 2014) 132}, [\href{http://arxiv.org/abs/1405.6814}{{\tt 1405.6814}}].

\bibitem{Deng_2006ApJ...643..426D}
L.~{Deng}, D.~R. {Xiong} and K.~L. {Chan}, \emph{{An Anisotropic Nonlocal
  Convection Theory}}, \href{http://dx.doi.org/10.1086/502707}{\emph{\apj} {\bf
  643} (May, 2006) 426--437}.

\bibitem{Xiong:2021FrASS...7...95X}
D.-r. {Xiong}, \emph{{Convection Theory and Relevant Problems in Stellar
  Structure, Evolution and Pulsation Stability Part I. Convection Theory and
  Structure of convection zone and stellar evolution}},
  \href{http://dx.doi.org/10.3389/fspas.2020.438864}{\emph{Frontiers in
  Astronomy and Space Sciences} {\bf 7} (May, 2021) 95}.

\bibitem{Stein_1998ApJ...499..914S}
R.~F. {Stein} and {\r{A}}.~{Nordlund}, \emph{{Simulations of Solar Granulation.
  I. General Properties}}, \href{http://dx.doi.org/10.1086/305678}{\emph{Apj}
  {\bf 499} (May, 1998) 914--933}.

\bibitem{Hotta_2023SSRv..219...77H}
H.~{Hotta}, Y.~{Bekki}, L.~{Gizon}, Q.~{Noraz} and M.~{Rast}, \emph{{Dynamics
  of Large-Scale Solar Flows}},
  \href{http://dx.doi.org/10.1007/s11214-023-01021-6}{\emph{Space~Sci.~Rev.}
  {\bf 219} (Dec., 2023) 77}, [\href{http://arxiv.org/abs/2307.06481}{{\tt
  2307.06481}}].

\bibitem{Miesch:2012ApJ...757..128M}
M.~S. {Miesch}, N.~A. {Featherstone}, M.~{Rempel} and R.~{Trampedach},
  \emph{{On the Amplitude of Convective Velocities in the Deep Solar
  Interior}}, \href{http://dx.doi.org/10.1088/0004-637X/757/2/128}{\emph{\apj}
  {\bf 757} (Oct., 2012) 128}, [\href{http://arxiv.org/abs/1205.1530}{{\tt
  1205.1530}}].

\bibitem{Lipari:2019jmk}
P.~Lipari and S.~Vernetto, \emph{{The shape of the cosmic ray proton
  spectrum}},
  \href{http://dx.doi.org/10.1016/j.astropartphys.2020.102441}{\emph{Astropart.
  Phys.} {\bf 120} (2020) 102441}, [\href{http://arxiv.org/abs/1911.01311}{{\tt
  1911.01311}}].

\bibitem{Mori:2009te}
M.~Mori, \emph{{Nuclear enhancement factor in calculation of Galactic diffuse
  gamma-rays: a new estimate with DPMJET-3}},
  \href{http://dx.doi.org/10.1016/j.astropartphys.2009.03.004}{\emph{Astropart.
  Phys.} {\bf 31} (2009) 341--343}, [\href{http://arxiv.org/abs/0903.3260}{{\tt
  0903.3260}}].

\bibitem{Kachelriess:2014mga}
M.~Kachelriess, I.~V. Moskalenko and S.~S. Ostapchenko, \emph{{Nuclear
  enhancement of the photon yield in cosmic ray interactions}},
  \href{http://dx.doi.org/10.1088/0004-637X/789/2/136}{\emph{Astrophys. J.}
  {\bf 789} (2014) 136}, [\href{http://arxiv.org/abs/1406.0035}{{\tt
  1406.0035}}].

\bibitem{Amenomori:2013own}
{\scshape Tibet ASgamma} collaboration, M.~Amenomori et~al., \emph{{Probe of
  the Solar Magnetic Field Using the ``Cosmic-Ray Shadow'' of the Sun}},
  \href{http://dx.doi.org/10.1103/PhysRevLett.111.011101}{\emph{Phys. Rev.
  Lett.} {\bf 111} (2013) 011101}, [\href{http://arxiv.org/abs/1306.3009}{{\tt
  1306.3009}}].

\bibitem{Tibet:2018PhRvL.120c1101A}
M.~{Amenomori}, X.~J. {Bi}, D.~{Chen}, T.~L. {Chen}, W.~Y. {Chen}, S.~W. {Cui}
  et~al., \emph{{Evaluation of the Interplanetary Magnetic Field Strength Using
  the Cosmic-Ray Shadow of the Sun}},
  \href{http://dx.doi.org/10.1103/PhysRevLett.120.031101}{\emph{\prl} {\bf 120}
  (Jan, 2018) 031101}, [\href{http://arxiv.org/abs/1801.06942}{{\tt
  1801.06942}}].

\bibitem{sunshadowAartsen:2018zrl}
{\scshape IceCube} collaboration, M.~Aartsen et~al., \emph{{Detection of the
  Temporal Variation of the Sun's Cosmic Ray Shadow with the IceCube
  Detector}},
  \href{http://dx.doi.org/10.3847/1538-4357/aaffd1}{\emph{Astrophys. J.} {\bf
  872} (2019) 133}, [\href{http://arxiv.org/abs/1811.02015}{{\tt 1811.02015}}].

\bibitem{sunshadowBeckerTjus:2019rqu}
J.~Becker~Tjus, P.~Desiati, N.~D{\"o}pper, H.~Fichtner, J.~Kleimann, M.~Kroll
  et~al., \emph{{Cosmic-Ray Propagation Around the Sun - Investigating the
  Influence of the Solar Magnetic Field on the Cosmic-Ray Sun Shadow}},
  \href{http://dx.doi.org/10.1051/0004-6361/201936306}{\emph{Astron.
  Astrophys.} {\bf 633} (2020) A83},
  [\href{http://arxiv.org/abs/1903.12638}{{\tt 1903.12638}}].

\bibitem{sunshadowAartsen:2020hzn}
{\scshape IceCube} collaboration, M.~G. Aartsen et~al., \emph{{Measurements of
  the time-dependent cosmic-ray Sun shadow with seven years of IceCube data:
  Comparison with the Solar cycle and magnetic field models}},
  \href{http://dx.doi.org/10.1103/PhysRevD.103.042005}{\emph{Phys. Rev. D} {\bf
  103} (2021) 042005}, [\href{http://arxiv.org/abs/2006.16298}{{\tt
  2006.16298}}].

\bibitem{Ingelman:1996mj}
G.~Ingelman and M.~Thunman, \emph{{High-energy neutrino production by cosmic
  ray interactions in the sun}},
  \href{http://dx.doi.org/10.1103/PhysRevD.54.4385}{\emph{Phys. Rev.} {\bf D54}
  (1996) 4385--4392}, [\href{http://arxiv.org/abs/hep-ph/9604288}{{\tt
  hep-ph/9604288}}].

\bibitem{Ng:2017aur}
K.~C.~Y. Ng, J.~F. Beacom, A.~H.~G. Peter and C.~Rott, \emph{{Solar Atmospheric
  Neutrinos: A New Neutrino Floor for Dark Matter Searches}},
  \href{http://dx.doi.org/10.1103/PhysRevD.96.103006}{\emph{Phys. Rev. D} {\bf
  96} (2017) 103006}, [\href{http://arxiv.org/abs/1703.10280}{{\tt
  1703.10280}}].

\bibitem{Arguelles:2017eao}
C.~A. Arg{\"u}elles, G.~de~Wasseige, A.~Fedynitch and B.~J.~P. Jones,
  \emph{{Solar Atmospheric Neutrinos and the Sensitivity Floor for Solar Dark
  Matter Annihilation Searches}},
  \href{http://dx.doi.org/10.1088/1475-7516/2017/07/024}{\emph{JCAP} {\bf 1707}
  (2017) 024}, [\href{http://arxiv.org/abs/1703.07798}{{\tt 1703.07798}}].

\bibitem{Edsjo:2017kjk}
J.~Edsjo, J.~Elevant, R.~Enberg and C.~Niblaeus, \emph{{Neutrinos from cosmic
  ray interactions in the Sun}},
  \href{http://dx.doi.org/10.1088/1475-7516/2017/06/033}{\emph{JCAP} {\bf 1706}
  (2017) 033}, [\href{http://arxiv.org/abs/1704.02892}{{\tt 1704.02892}}].

\bibitem{Aartsen:2019avh_solaratmneutrino}
{\scshape IceCube} collaboration, M.~Aartsen et~al., \emph{{Searches for
  neutrinos from cosmic-ray interactions in the Sun using seven years of
  IceCube data}},  \href{http://arxiv.org/abs/1912.13135}{{\tt 1912.13135}}.

\bibitem{IceCube:2021koo_icrcsolar}
{\scshape IceCube} collaboration, J.~Villarreal et~al., \emph{{Recent Progress
  in Solar Atmospheric Neutrino Searches with IceCube}},
  \href{http://dx.doi.org/10.22323/1.395.1174}{\emph{PoS} {\bf ICRC2021} (2021)
  1174}, [\href{http://arxiv.org/abs/2107.13696}{{\tt 2107.13696}}].

\bibitem{Aartsen:2016zhm_darkmatter}
{\scshape IceCube} collaboration, M.~Aartsen et~al., \emph{{Search for
  annihilating dark matter in the Sun with 3 years of IceCube data}},
  \href{http://dx.doi.org/10.1140/epjc/s10052-017-4689-9}{\emph{Eur. Phys. J.
  C} {\bf 77} (2017) 146}, [\href{http://arxiv.org/abs/1612.05949}{{\tt
  1612.05949}}].

\bibitem{Leane:2017vag}
R.~K. Leane, K.~C.~Y. Ng and J.~F. Beacom, \emph{{Powerful Solar Signatures of
  Long-Lived Dark Mediators}},
  \href{http://dx.doi.org/10.1103/PhysRevD.95.123016}{\emph{Phys. Rev. D} {\bf
  95} (2017) 123016}, [\href{http://arxiv.org/abs/1703.04629}{{\tt
  1703.04629}}].

\bibitem{Niblaeus:2019gjk}
C.~Niblaeus, A.~Beniwal and J.~Edsjo, \emph{{Neutrinos and gamma rays from
  long-lived mediator decays in the Sun}},
  \href{http://dx.doi.org/10.1088/1475-7516/2019/11/011}{\emph{JCAP} {\bf 11}
  (2019) 011}, [\href{http://arxiv.org/abs/1903.11363}{{\tt 1903.11363}}].

\bibitem{SWGO:2019ahw}
P.~Abreu et~al., \emph{{The Southern Wide-Field Gamma-Ray Observatory (SWGO): A
  Next-Generation Ground-Based Survey Instrument for VHE Gamma-Ray Astronomy}},
   \href{http://arxiv.org/abs/1907.07737}{{\tt 1907.07737}}.

\end{thebibliography}\endgroup

\end{document}